\documentclass[twocolumn,aps,prc,superscriptaddress,showpacs,floatfix]{revtex4}
\usepackage{graphicx}

\begin{document}
\title{Effect of three-body elastic scattering on heavy quark 
momentum degradation in the quark-gluon plasma}

\author{W. Liu}
\affiliation{Cyclotron Institute and Physics Department, Texas A$\&$M 
University, College Station, Texas 77843-3366}
\author{C. M. Ko} 
\affiliation{Cyclotron Institute and Physics Department, Texas A$\&$M 
University, College Station, Texas 77843-3366}

\begin{abstract}
Heavy quark drag coefficients in the quark-gluon plasma are evaluated 
in the perturbative QCD. It is found that for charm quarks the contribution 
from three-body elastic scattering is comparable to those from two-body 
elastic and radiative scatterings while for bottom quarks it becomes 
dominant. Using a schematic expanding fireball model, effects of both 
two-body and three-body scatterings on the transverse momentum spectra 
of heavy quarks produced in Au+Au collisions at center of mass energy 
$\sqrt{s_{NN}}=200$ GeV are studied. Results on electrons from 
resulting heavy meson decays are compared with available experimental data. 
\end{abstract}

 
\maketitle

One of the most interesting observations in central heavy ion collisions 
at the Relativistic Heavy Ion Collider (RHIC) is the suppressed production 
of hadrons with large transverse momentum ($p_T$)\cite{adcox,adler1}. This 
phenomenon has been attributed to the radiative energy loss of partonic 
jets produced from initial hard scattering of incoming nucleons as they 
pass through the created dense partonic matter  
\cite{wang,gyulassy,wiedemann}. The same mechanism fails, however, 
to explain a similarly large suppression of high $p_T$ charmed mesons 
observed through their decay electrons as a result of the dead cone 
effect associated with massive quarks \cite{dokshitzer,djordjevic1}. 
Furthermore, experimental data have indicated that charm quarks develop 
a substantial elliptic flow in non-central heavy ion collisions at 
RHIC \cite{adler}, consistent with the prediction of the quark coalescence 
model that assumes a thermally equilibrated charm quark distribution in 
the quark-gluon plasma (QGP) \cite{greco}. In both the Fokker-Planck 
approach \cite{moore} and the transport model \cite{molnar,zhang}, to 
reproduce the observed strong suppression of charm production at 
high $p_T$ and large charm elliptic flow requires a much larger charm 
quark two-body elastic scattering cross section than that given by the 
perturbative QCD. Such a large cross section would result if colorless 
resonances are formed in charm quark scattering with light quark 
\cite{hees}. On the other hand, it was recently realized \cite{wicks} 
that two-body elastic scattering with the pQCD cross section causes a 
similar energy loss for the charm quark as the two-body radiative 
scattering \cite{thoma}, and a large fraction of the observed suppression 
of charmed meson production at high $p_T$ can be accounted for when 
both effects are included. 

Since the density of the partonic matter formed in heavy ion collisions 
at RHIC is large, ranging from about 1 fm$^{-3}$ near hadronization 
to more than 10 fm$^{-3}$ during the initial stage, three-body elastic 
scattering may also contribute to charm quark energy loss in the QGP.
Previous studies have shown that gluon \cite{xu1} and quark \cite{xu2} 
three-body elastic scatterings are more efficient than two-body elastic 
scattering for the thermalization of initially produced partons.  In 
the present paper, the effect of three-body elastic scattering on heavy 
quark ($Q=c,b$) momentum degradation in the QGP is studied. 

\begin{figure}[ht]
\includegraphics[width=2.5in,height=1in,angle=0]{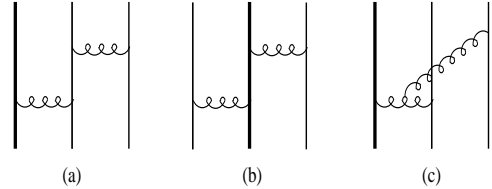} 
\caption{Topologically different diagrams for heavy quark (thick line) 
three-body elastic scatterings by quarks and antiquarks (thin line)
with different flavors. Wiggle lines denote the gluon.}
\label{diagram}
\end{figure}

For the processes $Qqq\to Qqq$, $Q\bar q\bar q\to Q\bar q\bar q$, and 
$Qq\bar q\to Qq\bar q$ with different light quark ($q$) and antiquark 
($\bar q$) flavors, there are three topologically different diagrams
in the lowest-order QCD as shown in Fig. \ref{diagram}. While diagram 
(c) corresponds to one diagram, two diagrams are generated from diagram 
(b), corresponding to either the left or the right gluon (wiggle line) 
is first exchanged, and four diagrams are generated from diagram (a) 
by exchanging the two gluons in all possible ways but keeping only one 
attached to the heavy quark.

To ensure true three-body scattering, the intermediate quark in both
diagrams (a) and (b) of Fig. \ref{diagram} must be off-shell, and this 
can be achieved by including its collisional width in the QGP and 
keeping only the real part of its propagator in evaluating these diagrams
as in the treatment of three-body scattering in hadronic matter 
\cite{batko}.  The quark width in QGP is given by 
$\Gamma/\hbar=\sum_i\langle\overline{|{\cal M}_i|^2}\rangle$, where the 
sum is over all scattering processes with $\overline{|{\cal M}_i|^2}$ 
being their squared amplitudes after averaging over the spins and colors 
of initial partons and summing over those of final partons.  The symbol 
$\langle\cdots\rangle$ denotes average over the thermal distributions of 
scattered quarks and antiquarks in the QGP and integration over the 
momenta of all final-state partons.

We first consider the quark width due to two-body elastic and radiative 
scatterings as well as the inverse process of the latter. For heavy quark 
two-body elastic scattering, there is only one $t$-channel gluon-exchange 
diagram for the process $Qq\to Qq$ or $Q\bar q\to Q\bar q$, while there 
are in addition one $s$-channel heavy quark pole and one $u$-channel heavy 
quark exchange diagram for the process $Qg\to Qg$. Diagrams for heavy quark
two-body radiative scattering are then obtained by adding an external 
gluon to above diagrams, leading to 5 diagrams for the process 
$Qq\to Qqg$ or $Q\bar q\to Q\bar qg$ and 16 diagrams for the process 
$Qg\to Qgg$. For the light quark, there are additional diagrams besides 
those similar to the ones for heavy quarks, and all these diagrams are 
included in our calculations. In evaluating these diagrams, 
we remove the collinear singularity in $t$-channel diagrams by using 
the screening mass $m_D=gT$ \cite{blazoit} for the exchanged gluon, where 
$g$ is the QCD coupling constant and $T$ is the temperature of the QGP, 
and the infrared singularity in two-body radiative scattering by 
including the thermal mass $m_g=m_D/\sqrt{2}$ \cite{blazoit} for the 
radiated gluon.   

Using the QCD coupling $\alpha_s=g^2/4\pi=0.3$, appropriate for the energy 
scales considered here, and also including the thermal mass for time-like
gluons as well as the thermal mass $m_q=m_D/\sqrt{6}$ \cite{blazoit} for 
time-like light quarks, we find that both heavy and light quark 
collisional widths in the QGP increase almost linearly with quark 
momentum as well as the temperature of the QGP.  For a momentum of  
4 GeV/$c$, the widths are about 75 MeV for charm quark and 49 MeV for 
bottom quark at $T=175$ MeV and increase, respectively, to about 130 MeV 
and 93 MeV at $T=350$ MeV. These widths are mainly due to two-body elastic 
scattering with only about 25\% from two-body radiative scattering and 
its inverse process. For the light quark, its width is about 50\% larger 
than that of charm quark. Using above calculated quark widths in the 
quark propagator, we find that three-body elastic scattering by quarks 
and antiquarks with different flavors increases quark widths by at 
most 10\% and can thus be neglected. We note that since the heavy quark 
width is smaller than the gluon thermal mass ($\Gamma_Q< m_g$), i.e., 
the time between heavy quark collisions is longer than the time for 
radiating a thermal gluon, the destructive Landau-Pomeranchuck-Migdal 
(LPM) interference effect \cite{djordjevic2}, which is neglected in 
present study, is expected to be small for heavy quark two-body 
radiative scattering as shown in Ref.\cite{djordjevic2}.

\begin{figure}[ht]
\vspace{0.5cm}
\includegraphics[width=2.8in,height=3.3in,angle=-90]{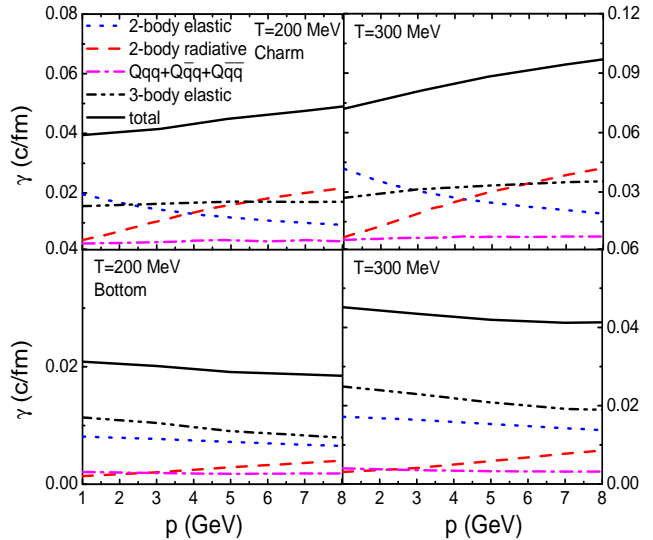}
\caption{Charm (upper panels) and bottom (lower panels) quark drag 
coefficients as functions of their momentum in a QGP of temperature 
$T=200$ MeV (left panels) or $T=300$ MeV (right panels).} 
\label{drag}
\end{figure} 

In the Fokker-Planck approach, the momentum degradation of a heavy quark 
in the QGP depends on its drag coefficient, which is given by averages 
similar to that for the quark collisional width, i.e., 
$\gamma(|{\bf p}|,T)=\langle\overline{|M|^2}\rangle 
-\langle\overline{|M|^2}{\bf p}\cdot{\bf p^\prime}\rangle/|\bf p|^2$
\cite{svetitsky,hees}. In the above, ${\bf p}$ and ${\bf p}^\prime$ 
are, respectively, the momenta of the heavy quark before and after 
a collision. In Fig. \ref{drag}, we show the charm (upper panels) 
and bottom (lower panels) quark drag coefficients as functions of their 
momentum in QGP at temperatures $T=200$ MeV (left panels) and $T=300$ MeV 
(right panels). It is seen that for charm quarks the contribution from 
three-body scattering by light quarks and antiquarks with different 
flavors (dash-dotted line) is generally smaller than those from 
two-body elastic (dotted line) and radiative scatterings (dashed line). 
For bottom quarks, the most important contribution is from 
two-body elastic scattering while two-body radiative and three-body 
elastic scattering give smaller but comparable contributions. We note 
that the contribution from diagram (b) of Fig. \ref{diagram} is larger
than that from diagram (a) by about one order of magnitude and that from 
diagram (c) by about two orders of magnitude.

If the two light quarks or antiquarks in Fig. \ref{diagram} have
same flavor, the number of diagrams is then doubled from interchanging 
final two light quarks or antiquarks. These exchange diagrams give 
same contribution as that due to the direct diagrams in 
Fig. \ref{diagram}. For the process $Qq\bar q\to Qq\bar q$ with same 
quark and antiquark flavor, besides diagrams similar to those of 
Fig. \ref{diagram}, the quark and antiquark can annihilate to a 
time-like virtual gluon, giving rise to additional five diagrams 
corresponding to different ways a gluon is exchanged between the 
heavy quark and the light quark, antiquark, and time-like gluon. The 
contribution from these annihilation diagrams is found to be small, 
and this allows us to neglect the interference between these diagrams 
and the diagrams in Fig. \ref{diagram}. Neglecting also the small
interference terms between above direct and exchange diagrams 
for scattering by two identical quarks or antiquarks, the contribution 
from quarks and antiquarks with same flavor is then the same as that 
from quarks and antiquarks with different flavors.

For heavy quark three-body elastic scattering involving gluons, there are
36 diagrams for the process $Qqg\to Qqg$ or $Q\bar qg\to Q\bar qg$, 
and 123 diagrams for the process $Qgg\to Qgg$. We have not been able to 
evaluate all these diagrams. Instead, we assume that these processes are 
also dominated by diagrams similar to diagram (b) of Fig. \ref{diagram}. 
This is also partially supported by the finding that for the process 
$qgg\to qgg$ involving a light quark the result obtained from including 
thermal and screening masses in the Parke formula \cite{parke} that takes 
into account many diagrams with topologies different from that of diagram 
(b) of Fig. \ref{diagram} indeed gives a much smaller contribution to 
the quark drag coefficient. Within this approximation, the contribution 
from $Qgg\to Qgg$ to heavy quark drag coefficients is slightly smaller 
while that from $Qqg\to Qqg$ and $Q\bar qg\to Q\bar qg$ is about a 
factor of three larger than that due to scattering by quarks and 
antiquarks, essentially due to the associated color and flavor factors 
in these processes. Because of the importance of $Qqg\to Qqg$ and 
$Q\bar qg\to Q\bar qg$, the heavy quark width is significantly increased 
by three-body elastic scattering.  Including this additional width in 
the intermediate heavy quark propagator in diagram (b) of Fig. 
\ref{diagram} leads to a charm quark drag coefficient due 
to three-body elastic scattering (dash-dot-dotted line in Fig. \ref{drag}) 
that is comparable to that due to two-body elastic or radiative scattering,
and a bottom quark drag coefficient that is now dominated by three-body
elastic scattering. The resulting total drag coefficients for heavy quarks 
due to both elastic two- and three-body scattering are shown by  
solid lines in Fig. \ref{drag}. 

To see the effect of three-body elastic scattering on heavy 
quark momentum degradation in QGP, we consider central Au+Au collisions 
at center-of-mass energy $\sqrt{s_{NN}}=200$ GeV. The initial $p_T$ 
spectra of charm and bottom quarks at midrapidity are taken to be 
${dN_c}/{d^2 p_T}=19.2[1+(p_T/6)^2]/\{(1+p_T/3.7)^{12}[1+\exp(0.9-2p_T)]\}$ 
and ${dN_b}/{d^2 p_T}=0.0025\left[1+(p_T/16)^5\right]\exp(-p_T/1.495)$,
respectively, with $p_T$ in unit of GeV/$c$.  Both are obtained by 
multiplying the heavy quark $p_T$ spectra from p+p collisions
at same energy by the number of binary collisions ($\sim 960$) in 
Au+Au collisions. For bottom quarks, their $p_T$ spectrum 
in p+p collisions is taken from the pQCD prediction of Ref.\cite{vogt}
as it is expected to be more reliable, while for charm quarks 
it is determined instead from fitting simultaneously measured $p_T$ 
spectrum of charmed mesons from d+Au collisions \cite{adams} and of 
electrons from heavy meson decays in p+p collisions. In obtaining the 
latter, heavy quarks are fragmented to hadrons via the Peterson 
fragmentation function $D(z)=1/\{z[1-1/z-\epsilon/(1-z)]^2\}$ \cite{gavai}, 
where $z$ is the fraction of heavy quark momentum carried by the formed
meson, with $\epsilon$ taken to be 0.02 for charm quarks and 0.002 
for bottom quarks in order to reproduce the empirical fragmentation 
functions used in Ref.\cite{vogt}. These heavy quarks are initially
distributed in the transverse plane according to that of the binary 
collision number, and their transverse momenta are directed isotropically 
in the transverse plane.

For the dynamics of formed QGP, we assume that it evolves boost invariantly 
in the longitudinal direction but with an accelerated transverse expansion. 
Specifically, its volume expands in the proper time $\tau$ according to 
$V(\tau)=\pi R(\tau)^2\tau$, where $R(\tau)=R_0+a/2(\tau-\tau_0)^2$
is the transverse radius with an initial value $R_0$=7 fm, 
$\tau_0$=0.6 fm is the QGP formation time, and $a=0.1c^2$/fm is the 
transverse acceleration \cite{chen}. With an initial temperature 
$T_i=350$ MeV and using thermal masses for quarks and gluons, this model 
gives a total transverse energy comparable to that measured in experiments. 
The time dependence of the temperature is then obtained from entropy 
conservation, and the critical temperature $T_c=175$ MeV is reached
at proper time $\tau_c\sim 5$ fm. 

For a heavy quark with an initial transverse momentum ${\bf p}_{0}$,
time evolution of its mean transverse momentum $\langle p_T\rangle$
can be obtained from the Fokker-Planck equation, i.e., 
$d\langle p_T\rangle/dt=-\langle\gamma(p_T,T)p_T\rangle$. Parametrizing 
the momentum dependence of heavy quark drag coefficients shown in 
Fig. \ref{drag} by $\gamma(p_T,T)\approx\gamma_0(T)[1+ap_T]$, we then 
have $d\langle p_T\rangle/dt\approx -\gamma_0(\langle p_T\rangle
+a\langle p_T^2\rangle)\approx -\gamma_0(\langle p_T\rangle
+a\langle p_T\rangle^2)$ if we take 
$\langle p_T^2\rangle\approx \langle p_T\rangle^2$, which
is valid for high transverse momentum heavy quarks as considered here. 
The final mean transverse momentum of the heavy quark after passing 
through the expanding QGP is then given by $\langle p_T\rangle=B/(1-aB)$, 
where $B=p_0\exp(-\int_{\tau_0}^{\tau_f}\gamma_0(\tau)d\tau)
/(1+ap_0)$ with $\tau_f$ denoting the smaller of the time when the QGP 
phase ends and the time for the heavy quark to escape the expanding QGP,
which depends on its initial transverse momentum and position. The 
final $p_T$ spectra of heavy quarks are then obtained 
by averaging over their initial spatial and transverse momentum 
distributions.

Besides fragmenting to mesons, heavy quarks produced in heavy ion 
collisions can also coalesce or recombine with thermal quarks in the 
QGP to form heavy mesons \cite{greco,zhang}. In both hadronization 
mechanisms, the momentum spectra of formed mesons are softer than those 
of heavy quarks. Instead of including both contributions, we use the 
fragmentation model in the present study for simplicity. Because of 
their smaller scattering cross sections with hadrons \cite{lin} and 
lower hadronic matter density, momentum degradation of heavy mesons in 
subsequent hadronic matter is small \cite{zhang} and is thus neglected. 
The heavy mesons produced from heavy quark fragmentation are therefore 
allowed to decay directly to electrons in order to compare with those 
measured in experiments. 

\begin{figure}[ht]
\includegraphics[width=2in,height=3.4in,angle=-90]{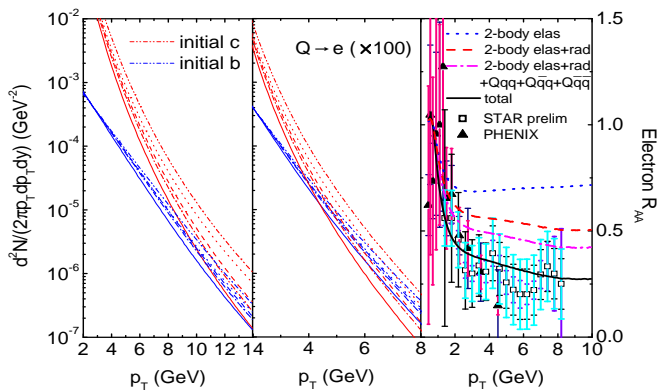} 
\caption{Initial and final transverse momentum spectra of heavy quarks 
(left panel) and their decay electrons (middle panel) as well as the
nuclear modification factor $R_{AA}$ for electrons (right panel) in 
central Au+Au collisions at $\sqrt{s_{NN}}=200$ GeV.}
\label{raa}
\end{figure}

In the left panel of Fig. \ref{raa}, we show our results for the 
initial (dash-dot-dotted lines) and final $p_T$ spectra 
of heavy quarks in Au+Au collisions at $\sqrt{s_{NN}}=200$ GeV. For both 
charm and bottom quarks, their final $p_T$ spectra become 
softer with the inclusion of more scattering processes: dotted lines
for two-body elastic scattering only, dashed lines for both two-body 
elastic and radiative scatterings, dash-dotted lines for adding also
three-body scattering by light quarks and antiquarks with different flavors, 
and solid lines for further including other three-body scattering processes.
Although bottom quarks are negligible at low transverse momentum, they 
are important at high transverse momentum as a result of their smaller 
momentum degradation in QGP than charm quarks. Due to larger bottom quark 
mass, electrons from their decays become dominant at high $p_T$ as shown in 
the middle panel of Fig. \ref{raa}. The ratio of the electron $p_T$ 
spectrum from final heavy mesons to that from initially produced ones, 
defined as the electron nuclear modification factor $R_{AA}$, is shown 
in the right panel of Fig. \ref{raa}. It is seen that the effect from 
two-body elastic scattering (dotted line) is as important as that from 
two-body radiative scattering, similar to that of Ref.\cite{wicks}. 
The electron $R_{AA}$ including both contributions (dashed line) 
is, however, slightly above the experimental data from the PHENIX 
collaboration (filled triangles) \cite{adler2} and the preliminary 
data from the STAR collaboration (open squares) \cite{bielcik}. Adding 
the contribution from charm quark three-body elastic scattering leads 
to an electron $R_{AA}$ that is in reasonable agreement with the measured 
one as shown by the solid line. If we include only charm quark 
three-body scattering by light quarks and antiquarks of different 
flavors, which is more reliably computed in present study, the resulting 
electron $R_{AA}$ is shown by the dash-dotted line which barely lies on 
the upper error bars of experimental data. 

Our results indicate that heavy quark three-body elastic scattering 
is important for understanding the electron nuclear modification factor 
in heavy ion collisions at RHIC. The most important heavy quark 
three-body elastic scattering process involves scattering 
with a gluon and a light quark or antiquark in the QGP, and this 
contribution has been evaluated using the assumption that they are 
dominated by $t$-channel gluon-exchange diagrams similar to diagram 
(b) of Fig. \ref{diagram} for three-body scattering by quarks and 
antiquarks with different flavors. Although we have shown that this 
is a valid approximation for the latter process, its validity in heavy 
quark three-body elastic scattering involving gluons remains to be 
verified. If more accurate calculations indeed give as large a three-body
contribution as shown here, then scattering involving more than three 
particles may also need to be considered. To evaluate the latter contribution
poses, however, a major theoretical challenge. 

We wish to thank Hendrik van Hees, Ralf Rapp, and Bin Zhang for helpful 
discussions. This work was supported in part by the US National Science 
Foundation under Grant No. PHY-0457265 and the Welch Foundation under 
Grant No. A-1358.


\begin{thebibliography}{99} 

\bibitem{adcox}A. Adcox {\it et al.} [PHENIX Collaboration], Phys. Rev. 
Lett. {\bf 88}, 022301 (2002).
\bibitem{adler1}C. Adler {\it et al.} [STAR Collaboration], Phys. Rev. 
Lett. {\bf 89}, 202301 (2002); {\bf 90}, 082302 (2002).
\bibitem{wang}X.N. Wang, Phys. Lett. B {\bf 579}, 299 (2004).
\bibitem{gyulassy}M. Gyulassy, P. L\'evai, and I. Vitev, 
Phys. Rev. Lett. {\bf 85}, 5535 (2001). 
\bibitem{wiedemann}U.A. Wiedemann, Nucl. Phys. B {\bf 588}, 303 (2000). 
\bibitem{dokshitzer}Y.L. Dokshitzer and D.E. Kharzeev, 
Phys. Lett. B {\bf 519}, 199 (2001).
\bibitem{djordjevic1}M. Djordjevic, M. Gyulassy, and S. Wicks, 
Phys. Rev. Lett. {\bf 94}, 112301 (2005).
\bibitem{adler}S.S. Adler {\it et al.} [PHENIX Collaboration], Phys. Rev.
C {\bf 72}, 024901 (2005). 
\bibitem{greco}V. Greco, C.M. Ko, and R. Rapp, 
Phys. Lett. B {\bf 595}, 202 (2004).
\bibitem{moore}G.D. Moore and D. Teaney, 
Phys. Rev. C {\bf 71}, 064904 (2005).
\bibitem{molnar}D. Molnar, J. Phys. G {\bf 31}, S421 (2005).
\bibitem{zhang}B. Zhang, L.W. Chen, and C.M. Ko, 
Phys. Rev. C {\bf 72}, 024906 (2005).
\bibitem{hees}H. van Hees and R. Rapp, Phys. Rev. C {\bf 71}, 034907 (2005);
H. van Hees {\it et al.}, 
nucl-th/0508055.
\bibitem{wicks}S. Wicks {\it et al.}, 
nucl-th/0512076.
\bibitem{thoma}M.G. Mustafa {\it et al.}, 
Phys. Lett. B {\bf 428}, 234 (1998).
\bibitem{xu1}X.M. Xu {\it et al.}, 
Nucl. Phys. A {\bf 744}, 347 (2004).
\bibitem{xu2}X.M. Xu {\it et al.}, 
Phys. Lett. B {\bf 629}, 68 (2005). 
\bibitem{batko}G. Batko {\it et al.}, 
Nucl. Phys. A {\bf 536}, 786 (1992).
\bibitem{blazoit}J.P. Blaizot and E. Iancu, Phys. Rep. {\bf 359}, 355 (2002).
\bibitem{djordjevic2}M. Djordjevic and M. Gyulassy, 
Nucl. Phys. A {\bf 733}, 265 (2004).
\bibitem{svetitsky}B. Svetitsky, Phys. Rev. D {\bf 37}, 2484 (1988).
\bibitem{parke}M. Mangano and S.J. Parke, Phys. Rep. {\bf 200}, 301 (1991).
\bibitem{vogt}M. Cacciari {\it et al.},
Phys. Rev. Lett. {\bf 95}, 122001 (2005). 
\bibitem{adams}J. Adams {\it et al.} [STAR Collaboration], Phys. Rev. Lett.
{\bf 94}, 062301 (2005).
\bibitem{gavai}R.V. Gavai {\it et al.}, 
Int. J. Mod. Phys. A {\bf 10}, 2999 (1995). 
\bibitem{chen}L.W. Chen {\it et al.}, 
Phys. Lett. B {\bf 601}, 34 (2004).
\bibitem{lin}Z.W. Lin {\it et al.}, 
Nucl. Phys. A {\bf 689}, 965 (2001).
\bibitem{adler2}S.S. Adler [PHENIX Collaboration], nucl-ex/0510047.
\bibitem{bielcik}J. Bielcik [STAR Collaboration], nucl-ex/0511005.

\end{thebibliography}
\end{document}